\documentclass{jetpl}

\usepackage{graphicx}

\twocolumn
\lat

\begin{document}

    \title{Nonsinusoidal current--phase relation in SFS Josephson junctions}
   \rtitle{Nonsinusoidal current--phase relation in SFS Josephson junctions}
 \sodtitle{Nonsinusoidal current--phase relation in SFS Josephson junctions}

 \author{A.\,A.~Golubov$^{\,+}$\/\thanks{e-mails: a.golubov@tn.utwente.nl, mkupr@pn.sinp.msu.ru,
                                                  fominov@landau.ac.ru},
         M.\,Yu.~Kupriyanov$^{\,*\,1)}$, and
         Ya.\,V.~Fominov$^{\,\square\,+\,1)}$}
 \rauthor{A.\,A.~Golubov, M.\,Yu.~Kupriyanov, and Ya.\,V.~Fominov}
 \sodauthor{Golubov, Kupriyanov, and Fominov}

\address{
$^+$ Department of Applied Physics, University of Twente, 7500 AE Enschede, The Netherlands\\
$^*$ Nuclear Physics Institute, Moscow State University, 119899 Moscow, Russia\\
$^\square$ L.\,D.~Landau Institute for Theoretical Physics RAS, 117940 Moscow, Russia}

\dates{25 April 2002}{*}

\abstract{Various types of the current--phase relation $I(\varphi)$ in
superconductor--ferromagnet--superconductor (SFS) point contacts and planar double-barrier junctions are
studied within the quasiclassical theory in the limit of thin diffusive ferromagnetic interlayers. The
physical mechanisms leading to highly nontrivial $I(\varphi)$ dependence are identified by studying the
spectral supercurrent density. These mechanisms are also responsible for the $0$--$\pi$ transition in SFS
Josephson junctions.}


\maketitle

The relation between the supercurrent $I$ across a Josephson junction and the difference $\varphi$ between
the phases of the order parameters in the superconducting banks is an important characteristic of the
structure. The form of $I(\varphi)$ dependence is essentially used for analyzing the dynamics of systems
containing Josephson junctions \cite{LikharevK}. Studying $I(\varphi)$ also provides information on
pairing symmetry in superconductors \cite{Il'ichev}.

In structures with tunnel-type conductivity of a weak link (SIS) the current--phase relation is
sinusoidal, $I (\varphi)=I_c \sin\varphi$ with $I_c>0$, in the whole temperature range below the critical
temperature. At the same time, in point contacts (ScS) and junctions with metallic type of conductivity
(SNS) strong deviations from the sinusoidal form take place at low temperatures $T$ \cite{Likharev} with
the maximum of $I (\varphi)$ achieved at $\pi/2 < \varphi_\mathrm{max} < \pi$.

The situation drastically changes if there is magnetoactive material in the region of weak link. The
transition from the $0$-state [$I (\varphi)=I_c \sin\varphi$ with $I_c>0$] to the $\pi$-state ($I_c<0$) in
junctions containing ferromagnets has been theoretically predicted in a variety of Josephson structures
\cite{Bulaevski,Buzdin,Buzdin1,Yip,Wilhelm,Kriv,Chtch,Barash,GKF} and experimentally observed in SFS and
SIFS junctions \cite{Ryazanov,Kontos}. In the general case modifications of $I(\varphi)$ do not reduce to
the $0$--$\pi$ transition. It was shown that tunneling across ferromagnetic insulator (F$_\mathrm{I}$) in
clean SF$_\mathrm{I}$S junctions \cite{Tanaka} or across a magnetically active interface between two
superconductors \cite{Fogelstrom} may result in a nonsinusoidal shape of $I(\varphi)$ due to shift of
Andreev bound states. Similar situation occurs in long SFS junctions with ideally transparent interfaces
in the clean \cite{Dobro} and diffusive \cite{Yip,Wilhelm} regimes. However, in the latter case the
effects take place only in a narrow interval of very low temperatures (due to smallness of the Thouless
energy), while here we shall consider short-length structures where the effects are more pronounced and
exist practically in the whole temperature range (the role of temperature will be discussed elsewhere
\cite{future}).

In this letter we investigate anomalies of the $I(\varphi)$ relation in several types of SFS structures
which allow analytical solution while have not been fully explored yet: the SFcFS point contact with clean
or diffusive constriction as a weak link, and the double-barrier SIFIS junction; the ferromagnetic layers
are assumed to be thin, and the magnetization is homogeneous throughout the F part of the system. In
particular, we show that the maximum of $I(\varphi)$ can shift from $\pi/2 \leq \varphi_\mathrm{max} <
\pi$ to $0 < \varphi_\mathrm{max} < \pi /2$ as a function of the exchange field in the ferromagnet.
Previously, current--phase relation of this type was theoretically predicted either if superconductivity
in the S-electrodes was suppressed by the supercurrent in the SNS structure \cite{Meriakri,Zubkov,Kupr} or
in the vicinity of $T=0$ in long SFS junctions \cite{Yip,Wilhelm}.

The outline of the paper is as follows. We start with studying the SFcFS structure composed of two SF
sandwiches linked by a clean Sharvin constriction with arbitrary transparency $D$. We show that the
energy--phase relation of this junction can have \emph{two} minima: at $\varphi=0$ and $\varphi=\pi$
(while the energy of the junction in the pure $0$- or $\pi$-state has a \emph{single} minimum --- at
$\varphi=0$ or $\varphi=\pi$, respectively). As a result, $I(\varphi)$ dependence can intersect zero not
only at $\varphi=0$ and $\varphi =\pi$ but also at an arbitrary value $\varphi_0$ from the interval $0<
\varphi_0< \pi$. The salient effects which occur in junctions with clean constriction survive averaging
over the distribution of transmission eigenvalues and thus occur also in diffusive point contacts.
Physically, the properties of SFS structures are explained by splitting of Andreev levels due to the
exchange field; to demonstrate this, we study the spectral supercurrent. Finally, we show that the same
mechanism provides shifting of the $I(\varphi)$ maximum to $\varphi < \pi /2$ in the double-barrier SIFIS
junctions which can be more easily realized in experiment.

\textbf{SFcFS with clean constriction.} We start with a model structure composed of two superconducting SF
bilayers connected by a clean constriction with transparency $D$ (the size of the constriction $a$ is much
smaller than the mean free path $l$: $a\ll l$). We assume that the S-layers are bulk and that the dirty
limit conditions are fulfilled in the S- and F-metals. For simplicity we also assume that the parameters
of the SF interfaces $\gamma$ and $\gamma_B$ obey the condition
\begin{gather}
\gamma \ll \max (1,\gamma _B), \label{cond1} \\
\gamma_B=R_B\mathcal{A}_B/ \rho_F \xi_F,\quad \gamma =\rho_S \xi_S/ \rho_F \xi_F, \notag
\end{gather}
where $R_B$ and $\mathcal{A}_B$ are the resistance and the area of the SF interfaces; $\rho_{S(F)}$ is the
resistivity of the S (F) material, and the coherence lengths are related to the diffusion constants
$D_{S(F)}$ as $\xi_{S(F)}=\sqrt{D_{S(F)}/2\pi T_c}$, where $T_c$ is the critical temperature of the
S-material. We shall consider symmetric structure and restrict ourselves to the limit when the thickness
of the F-layers is small:
\begin{equation} \label{cond2}
d_F \ll \min \left( \xi_F, \sqrt{D_F/2H}\right),
\end{equation}
where $H$ is the exchange energy in the F-layers.

Under condition (\ref{cond1}), we can neglect the suppression of superconductivity in the S-electrodes by
the supercurrent and the proximity effect, and reduce the problem to solving the Usadel equations
\cite{Usadel} in the F-layers
\begin{equation}
\xi_F^2 \frac \partial {\partial x}\ \left[ G_F^2 \frac \partial {\partial x} \Phi_F \right]
-\frac{\widetilde\omega}{\pi T_c} G_F \Phi_F =0, \label{UsadelFi}
\end{equation}
with the boundary conditions at the SF interfaces ($x=\mp d_F$) in the form \cite{KL}
\begin{gather}
\pm \gamma_B \frac{\xi_F G_F}{\widetilde\omega} \frac \partial {\partial x} \Phi_F =G_S \left(
\frac{\Phi_F}{\widetilde\omega} -\frac{\Phi_S}\omega \right) , \\
G_S = \omega/ \sqrt{\omega^2+\Delta_0^2},\quad \Phi_S (\mp d_F)=\Delta_0 \exp \left( \mp i \varphi/2
\right). \notag
\end{gather}
In the above equations the $x$ axis is perpendicular to the interfaces with the origin at the
constriction; $\omega =\pi T(2n+1)$ are Matsubara frequencies; $\widetilde\omega =\omega +iH$; and
$\Delta_0$ is the absolute value of the pair potential in the superconductors. The function $\Phi$
parameterizes the Usadel functions $G$, $F$, and $\bar F$:
\begin{gather}
G_F(\omega) =\frac{\widetilde\omega}{\sqrt{\widetilde\omega^2 +\Phi_F(\omega) \Phi_F^*(-\omega)}},
\label{def_f}\\
F_F(\omega) =\frac{\Phi_F(\omega)}{\sqrt{\widetilde\omega^2+\Phi_F(\omega) \Phi_F^*(-\omega)}},\quad \bar
F_F(\omega) = F_F^*(-\omega). \notag
\end{gather}

Under condition (\ref{cond2}), the spatial gradients in the F-layers arising due to the proximity effect
and current are small. Then we can expand the solution of Eqs. (\ref{UsadelFi})--(\ref{def_f}) up to the
second order in small gradients, arriving at \cite{GKF}
\begin{equation} \label{Phi_0}
\Phi_{F1,F2} = \Phi_0 \exp(\mp i\varphi/2),\quad \Phi_0 = \Delta_0 \widetilde\omega / W,
\end{equation}
where
\begin{gather}
W=\omega + \widetilde\omega \gamma_{BM} \Omega,\qquad \Omega =\sqrt{\omega^2 +\Delta_0^2}/\pi T_c, \\
\gamma_{BM} =\gamma_B d_F/ \xi_F, \notag
\end{gather}
and the indices $1$ and $2$ refer to the left- and right-hand side of the constriction, respectively.

The supercurrent in constriction geometry is given by the general expression \cite{Zaitsev}
\begin{equation}
I=\frac{4\pi T}{e R_N} \Imag \sum_{\omega>0} \frac{(\bar F_1 F_2 -F_1 \bar F_2)/2}{2 - D \left[ 1 - G_1
G_2 - (\bar F_1 F_2 +F_1 \bar F_2)/2 \right] },
\end{equation}
where $R_N$ is the normal-state resistance of the junction. Inserting Eq. (\ref{Phi_0}) in this expression
we obtain
\begin{equation} \label{Constr_D}
I=\frac{2\pi T}{e R_N} \Real \sum_{\omega >0} \frac{\Delta_0^2 \sin\varphi}{W^2+\Delta_0^2 \left[ 1 -D
\sin^2(\varphi /2) \right]}.
\end{equation}
Finally, the current--phase relation takes the form
\begin{gather}
I(\varphi)=\frac{2\pi T}{e R_N}\sum_{\omega >0} \frac{A \Delta_0^2 \sin\varphi}{A^2+B^2}, \\
A = \Delta_0^2 \left[ 1-D\sin^2 \left( \varphi/2 \right) \right] - H^2 \left( \gamma_{BM} \Omega \right)^2
\notag \\
+\omega^2 \left( 1+\gamma_{BM} \Omega \right)^2 , \notag \\
B= 2 \omega H \gamma_{BM} \Omega \left( 1+\gamma_{BM} \Omega \right) . \notag
\end{gather}

\begin{figure}[tb]
 \includegraphics[width=84mm]{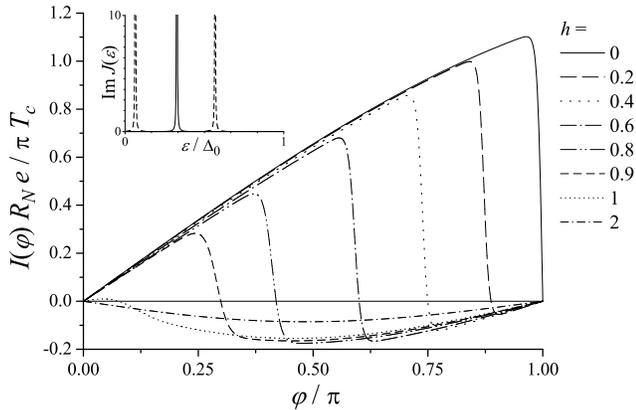}
\caption{Fig.\ref{fig:fig1}. Current--phase relation in clean SFcFS junction with ideally transparent
constriction ($D=1$) at $T/T_c=0.01$, $\gamma_{BM}=1$ for different values of the normalized exchange
field $h$. Inset: spectral supercurrent density at $\varphi=2 \pi /3$ for $h=0$ (solid line) and $h=0.4$
(dashed line).}
 \label{fig:fig1}
\end{figure}

\begin{figure}[tb]
 \includegraphics[width=84mm]{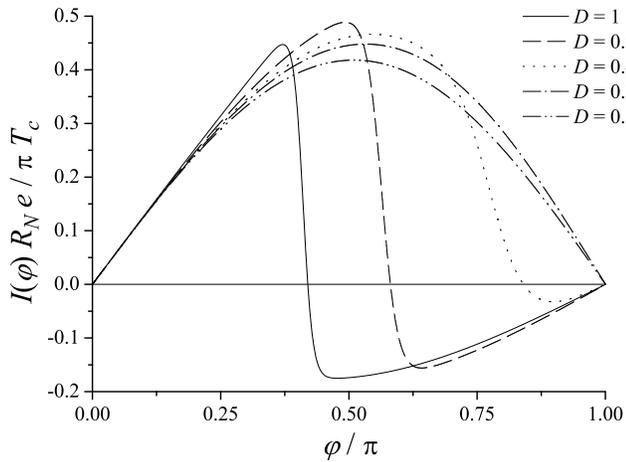}
\caption{Fig.\ref{fig:fig2}. Current--phase relation in clean SFcFS junction at $T/T_c=0.01$,
$\gamma_{BM}=1$, $h=0.8$ for different values of the barrier transparency $D$.}
 \label{fig:fig2}
\end{figure}

At small $\omega$ the function $A$ [and thus $I(\varphi)$] changes its sign at finite phase difference
$\varphi_c =2\arcsin \sqrt{[1-(\gamma_{BM} h)^2]/D}$ if the exchange field is in the range $1-D <
(\gamma_{BM} h)^2 <1$; here $h$ is the normalized exchange field, $h=H/\pi T_c$. The results for
$I(\varphi)$ are shown in Figs.\ref{fig:fig1},\ref{fig:fig2} and can be understood by considering the
spectral supercurrent density $\Imag J(\varepsilon)$. The latter is obtained by the analytical
continuation in Eq. (\ref{Constr_D}) and is given by a sum of delta-functions $\delta (\varepsilon-E_B)$
where $E_B$ are energies of the Andreev bound states. At $\gamma_{BM}=0$ the well-known result $E_B =\pm
\Delta_0 \sqrt{1-D \sin^2 (\varphi /2)}$ is reproduced, while at finite $\gamma_{BM}$ the exchange field
splits each bound state into two (see inset in Fig.\ref{fig:fig1}). At $\varphi=\varphi_c$ one of these
split (positive) peaks crosses zero leaving the domain $\varepsilon>0$, and simultaneously a negative peak
moves from the region $\varepsilon<0$ into $\varepsilon>0$ reversing the sign of the supercurrent.

The sign-reversal of the supercurrent (the $0$--$\pi$ transition) can also be achieved at \emph{fixed} $H$
due to nonequilibrium population of levels. This phenomenon has been studied in long diffusive SNS
\cite{Volkov,Wilhelm1,Basel} and SFS junctions \cite{Yip,Wilhelm}.

\textbf{SFcFS with diffusive constriction.} To get the $I(\varphi)$ relation for the diffusive point
contact [$l \ll a\ll \min( \xi_F, \sqrt{D_F/2H} )$] we integrate $\int_0^1 \rho(D) I(D) dD$, where $I(D)$
is given by Eq. (\ref{Constr_D}) for the clean case (note that $R_N \propto D^{-1}$ in this equation) and
$\rho(D)$ is Dorokhov's density function $\rho(D)= 1/ 2 D \sqrt{1-D}$ \cite{Dorokhov}. Finally, we arrive
at the result
\begin{align}
I(\varphi) = & \frac{4\pi T}{e R_N} \Real \sum_{\omega >0} \frac{\Delta_0 \cos (\varphi
/2)}{\sqrt{W^2+\Delta_0^2 \cos^2 (\varphi /2)}} \notag \\
& \times \arctan \left( \frac{\Delta_0 \sin (\varphi /2)}{\sqrt{W^2+\Delta_0^2 \cos^2 (\varphi /2)}}
\right).
\end{align}
This expression coincides with the direct solution of the Usadel equations, and at $\gamma_{BM}=0$ it
reproduces the Kulik--Omelyanchuk formula for the diffusive ScS constriction \cite{KO1}.

\begin{figure}[tb]
 \includegraphics[width=84mm]{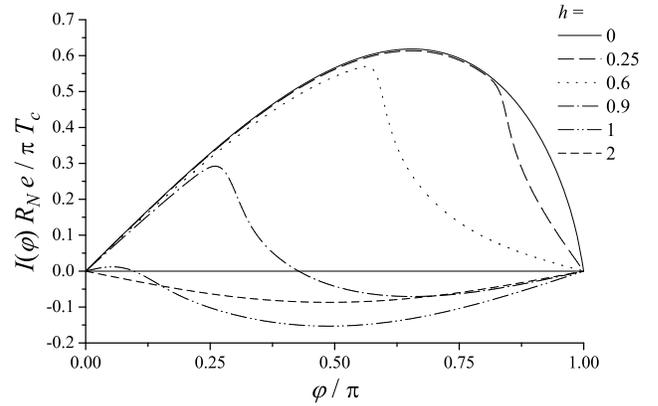}
\caption{Fig.\ref{fig:fig3}. Current--phase relation in diffusive SFcFS point contact at $T/T_c=0.01$,
$\gamma_{BM}=1$ for different values of the exchange field $h$.}
 \label{fig:fig3}
\end{figure}

Calculation of $I(\varphi)$ using the above expression yields results similar to those for the clean point
contact, however the transition from $0$- to $\pi$-state becomes less sharp (see Fig.\ref{fig:fig3}).

Temperature dependence of the critical current in this case shows thermally-induced $0$--$\pi$ crossover
with nonzero critical current at the transition point, in agreement with results of
Refs.~\cite{Wilhelm,Chtch} (results for $I_c(T)$ will be presented elsewhere \cite{future}). This is a
natural result since the barrier transparency is high and the current--phase relation is strongly
nonsinusoidal. We note that in Ref.~\cite{Ryazanov} the measured critical current vanished at the
$0$--$\pi$ transition point because of the low transparency regime (and hence sinusoidal current--phase
relation) realized in that experiment.

\textbf{SIFIS.} Now we turn to a double-barrier SIFIS junction (I denotes an insulating barrier)
--- this structure is easier for experimental implementation than an SFcFS junction. In the case of SIFIS,
due to dephasing effects (this situation is similar to the SINIS junction \cite{Brinkman}) the
supercurrent can not be derived by integrating over the corresponding transmission distribution (except
for the case of vanishing $\gamma_{BM}$) and must be calculated by solving the Usadel equations.

We assume that condition (\ref{cond1}) is satisfied; then we can neglect the suppression of
superconductivity in the S-electrodes by the supercurrent and the proximity effect. In this case the
system is described by Eqs. (\ref{UsadelFi})--(\ref{def_f}), although now instead of two F-layers
connected by a constriction we have a continuous F-layer (at $-d_F <x< d_F$).

We also assume that the F-layer is thin [condition (\ref{cond2})] and that $\gamma_B \gg d_F/\xi_F$, hence
the spatial gradients in the F-layer are small. Then (similarly to the case of constriction) we can expand
the solution of Eqs. (\ref{UsadelFi})--(\ref{def_f}) up to the second order in small gradients, arriving
at
\begin{gather}
\Phi_F =\Phi_0 \cos( \varphi/ 2 )
 + i\frac{\widetilde\omega G_S}{\omega G_F} \frac{\Delta_0 \sin (\varphi /2)}{\gamma_B} \frac x{\xi_F},
 \label{Sol1} \\
G_F=\frac{\widetilde\omega}{\sqrt{\widetilde\omega^2+\Phi_0^2 \cos^2 ( \varphi/ 2 )}}, \label{Sol1a}
\end{gather}
with $\Phi_0$ defined in Eq. (\ref{Phi_0}) [in the final result (\ref{Sol1}) we retained only the first
order in gradients --- this accuracy is sufficient for calculating the current].

\begin{figure}[tb]
 \includegraphics[width=84mm]{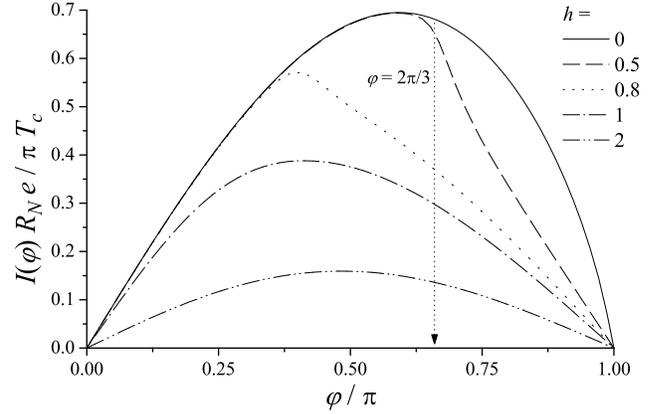}
\caption{Fig.\ref{fig:fig4}. Current--phase relation in double-barrier SIFIS junction at $T/T_c=0.02$,
$\gamma_{BM}=1$ for different values of the exchange field $h$. The value $\varphi=2\pi/3$ will be used in
Fig.\ref{fig:fig5}.}
 \label{fig:fig4}
\end{figure}

Inserting the solution (\ref{Sol1}), (\ref{Sol1a}) into the general expression for the supercurrent
\begin{equation}
I= -\frac{\pi T \mathcal{A}_B}{e \rho_F} \Imag \sum_\omega \frac{G_F^2(\omega)}{\widetilde\omega^2}
\Phi_F(\omega) \frac\partial{\partial x} \Phi_F^*(-\omega),
\end{equation}
we obtain
\begin{equation}
I(\varphi) =\frac{2\pi T}{e R_N} \Real \sum_{\omega >0} \frac{\Delta_0^2
\sin\varphi}{\sqrt{\omega^2+\Delta_0^2} \sqrt{W^2+\Delta_0^2 \cos^2(\varphi /2)}} \label{IotFi}
\end{equation}
(our assumptions imply that $R_N\approx 2 R_B$). This result demonstrates that the SIFIS junction with
thin F-layer is always in the $0$-state.\footnote{$\vphantom{\text{\Huge I}}$In the case under discussion
when the F-layers are thin and the interface parameters obey condition (\ref{cond1}), the phase of the
pair potential is constant in the S-part and almost constant in the F-part, however it jumps at the two SF
interfaces \cite{GKF}. The two jumps compensate each other in SIFIS with a single F-layer, whereas in
SFcFS they add up at the weak link thus opening possibility for the $\pi$-state.} Nevertheless,
$I(\varphi)$ is strongly modified by finite $H$ (see Fig.\ref{fig:fig4}), especially at small
temperatures. Figure~\ref{fig:fig4} clearly demonstrates that an increase of $H$ results not only in
suppression of the critical current, but also in the shift of the $I(\varphi)$ maximum from
$\varphi_\mathrm{max} \approx 1.86$ at $H=0$ to the values smaller than $\pi /2$. In the limit of large
exchange fields, $h \gg \gamma_{BM}^{-1}$, $I(\varphi)$ returns to the sinusoidal form.

\begin{figure}[tb]
 \includegraphics[width=84mm]{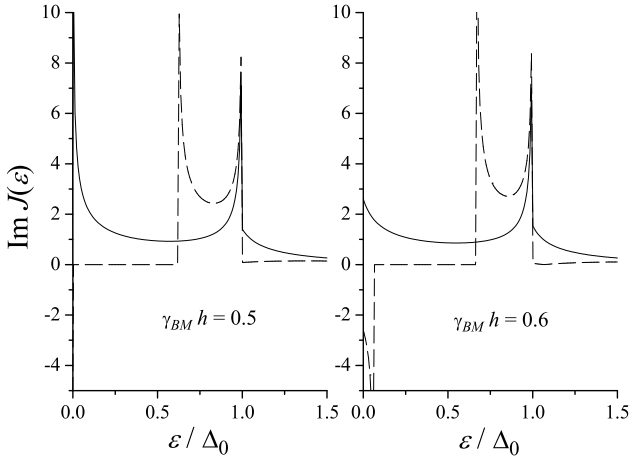}
\caption{Fig.\ref{fig:fig5}. Spectral supercurrent in diffusive double-barrier SIFIS junction with thin
ferromagnetic interlayer at $\gamma_{BM}=1$, $\varphi=2\pi/3$ for two values of the exchange field $h$.
The chosen value of $\varphi$ corresponds to $\varphi_c$ at $\gamma_{BM} h=0.5$, and the figure
demonstrates that the positive peak for one spin projection disappears from while the negative peak for
the other projection appears in the domain $\varepsilon>0$.}
 \label{fig:fig5}
\end{figure}

The physical origin of these results can be clarified in the real energy $\varepsilon$ representation.
Making analytical continuation in Eq. (\ref{IotFi}) by replacement $\omega \rightarrow
-i(\varepsilon+i0)$, we obtain the spectral supercurrent density $\Imag J(\varepsilon)$ which contains
contributions of Andreev bound states with different energies:
\begin{gather}
I=\frac 1{4 e R_N} \sum_{\sigma=\pm 1} \int \Imag J(\varepsilon, \sigma H) \tanh
\left( \frac \varepsilon {2T} \right) d\varepsilon, \\
\Imag J(\varepsilon,H)= \Imag \frac{\Delta_0^2 \sin\varphi}{\sqrt{\Delta_0^2 -\varepsilon^2}
\sqrt{\Delta_0^2 \cos^2 (\varphi /2)-\widetilde\varepsilon^2}}, \label{SpectrJ} \\
\widetilde\varepsilon =\varepsilon +\gamma_{BM} (\varepsilon -H) \Omega (\varepsilon),\quad \Omega
(\varepsilon)= \sqrt{\Delta_0^2-\varepsilon^2}/ \pi T_c. \notag
\end{gather}
Equation (\ref{SpectrJ}) implies that at $\varphi_c =2 \arccos (\gamma_{BM} h)$ singularities in $\Imag
J(\varepsilon)$ are shifted to the Fermi level. At $\varphi > \varphi_c$ the negative singularity in
$\Imag J(\varepsilon)$ for one spin projection crosses the Fermi level and appears in the positive energy
domain, whereas the positive peak for the other projection leaves the domain $\varepsilon>0$ (this process
is illustrated in Fig.\ref{fig:fig5}). As a result, the contribution to the supercurrent from low energies
changes its sign, and the supercurrent $I(\varphi)$ becomes suppressed at $\varphi > \varphi_c$ (see
Fig.\ref{fig:fig4}). However, at higher energies $\varepsilon \sim \Delta_0$ modifications in $\Imag
J(\varepsilon)$ are weak, and the resulting $I(\varphi)$ does not change its sign.

In conclusion, we have studied nonsinusoidal current--phase relation in Josephson junctions with thin
ferromagnetic interlayers and identified the physical mechanisms of these effects in terms of splitting of
Andreev bound states in the junction by the exchange field. In particular, we have shown that zero-energy
crossing of Andreev bound states is responsible for the sign-reversal of $I(\varphi)$, which also survives
averaging over distribution of transmission eigenvalues in the diffusive junction. As a result, the
energy--phase relation for the junction has two minima: at $\varphi=0$ and $\varphi=\pi$. The phenomena
studied in this work may be used for engineering cryoelectronic devices manipulating spin-polarized
electrons and in qubit circuits.

We acknowledge stimulating discussions with J.~Aarts, N.\,M.~Chtchelkatchev, K.\,B.~Efetov,
M.\,V.~Feigel'man, V.\,V.~Ryazanov, and M.~Siegel. The research of M.Yu.K. was supported by the Russian
Ministry of Industry, Science and Technology (RMIST). Ya.V.F. acknowledges financial support from the
Russian Foundation for Basic Research (project 01-02-17759), Forschungszentrum J\"ulich (Landau
Scholarship), the Swiss National Foundation, RMIST, and the program ``Quantum Macrophysics'' of the
Russian Academy of Sciences (RAS).


\begin{thebibliography}{99}

\bibitem{LikharevK}
K.\,K.~Likharev, \emph{Dynamics of Josephson Junctions and Circuits} (Gordon and Breach, Amsterdam, 1991).

\bibitem{Il'ichev}
E.~Il'ichev, M.~Grajcar, R.~Hlubina et al., Phys. Rev. Lett. \textbf{86}, 5369 (2001).

\bibitem{Likharev}
K.\,K.~Likharev, Rev. Mod. Phys. \textbf{51}, 101 (1979).

\bibitem{Bulaevski}
L.\,N.~Bulaevskii, V.\,V.~Kuzii, and A.\,A.~Sobyanin, Pis'ma Zh. Eksp. Teor. Fiz. \textbf{25}, 314 (1977)
[JETP Lett. \textbf{25}, 290 (1977)].

\bibitem{Buzdin}
A.\,I.~Buzdin, L.\,N.~Bulaevskii, and S.\,V.~Panyukov, Pis'ma Zh. Eksp. Teor. Fiz. \textbf{35}, 147 (1982)
[JETP Lett. \textbf{35}, 178 (1982)].

\bibitem{Buzdin1}
A.\,I.~Buzdin and M.\,Yu.~Kupriyanov, Pis'ma Zh. Eksp. Teor. Fiz. \textbf{53}, 308 (1991) [JETP Lett.
\textbf{53}, 321 (1991)].

\bibitem{Yip}
S.-K. Yip, Phys. Rev.~B \textbf{62}, R6127 (2000).

\bibitem{Wilhelm}
T.\,T.~Heikkil\"{a}, F.\,K.~Wilhelm, and G.~Sch\"{o}n, Europhys. Lett. \textbf{51}, 434 (2000).

\bibitem{Kriv}
E.\,A.~Koshina and V.\,N.~Krivoruchko, Phys. Rev.~B \textbf{63}, 224515 (2001).

\bibitem{Chtch}
N.\,M.~Chtchelkatchev, W.~Belzig, Yu.\,V.~Nazarov, and C.~Bruder, Pis'ma Zh. Eksp. Teor. Fiz. \textbf{74},
357 (2001) [JETP Lett. \textbf{74}, 323 (2001)].

\bibitem{Barash}
Yu.\,S.~Barash and I.\,V.~Bobkova, Phys. Rev.~B \textbf{65}, 144502 (2002).

\bibitem{GKF}
A.\,A.~Golubov, M.\,Yu.~Kupriyanov, and Ya.\,V.~Fominov, Pis'ma Zh. Eksp. Teor. Fiz. \textbf{75}, 223
(2002) [JETP Lett. \textbf{75}, 190 (2002)].

\bibitem{Ryazanov}
V.\,V.~Ryazanov, V.\,A.~Oboznov, A.\,Yu.~Rusanov et al., Phys. Rev. Lett. \textbf{86}, 2427 (2001).

\bibitem{Kontos}
T.~Kontos, M.~Aprili, J.~Lesueur et al., cond-mat/0201104.

\bibitem{Tanaka}
Y.~Tanaka and S.~Kashiwaya, Physica~C \textbf{274}, 357 (1997).

\bibitem{Fogelstrom}
M.~Fogelstr\"{o}m, Phys. Rev.~B \textbf{62}, 11812 (2000).

\bibitem{Dobro}
L.~Dobrosavljevi\'{c}--Gruji\'{c}, R.~Ziki\'{c}, and Z.~Radovi\'{c}, Physica C \textbf{331}, 254 (2000).

\bibitem{future}
A.\,A.~Golubov, M.\,Yu.~Kupriyanov, and Ya.\,V.~Fominov, in preparation.

\bibitem{Meriakri}
Z.\,G.~Ivanov, M.\,Yu.~Kupriyanov, K.\,K.~Likharev et al., Fiz. Nizk. Temp. \textbf{7}, 560 (1981) [Sov.
J. Low Temp. Phys. \textbf{7}, 274 (1981)].

\bibitem{Zubkov}
A.\,A.~Zubkov and M.\,Yu.~Kupriyanov, Fiz. Nizk. Temp. \textbf{9}, 548 (1983) [Sov. J. Low Temp. Phys.
\textbf{9}, 279 (1983)].

\bibitem{Kupr}
M.\,Yu.~Kupriyanov, Pis'ma Zh. Eksp. Teor. Fiz. \textbf{56}, 414 (1992) [JETP Lett. \textbf{56}, 399
(1992)].

\bibitem{Usadel}
K.\,D.~Usadel, Phys. Rev. Lett. \textbf{25}, 507 (1970).

\bibitem{KL}
M.\,Yu.~Kupriyanov and V.\,F.~Lukichev, Zh. Eksp. Teor. Fiz. \textbf{94}, 139 (1988) [Sov. Phys. JETP
\textbf{67}, 1163 (1988)].

\bibitem{Zaitsev}
A.\,V.~Zaitsev, Zh. Eksp. Teor. Fiz. \textbf{86}, 1742 (1984) [Sov. Phys. JETP \textbf{59}, 1015 (1984)].

\bibitem{Volkov}
A.\,F.~Volkov, Phys. Rev. Lett. \textbf{74}, 4730 (1995).

\bibitem{Wilhelm1}
F.\,K.~Wilhelm, G.~Sch\"{o}n, and A.\,D.~Zaikin, Phys. Rev. Lett. \textbf{81}, 1682 (1998).

\bibitem{Basel}
J.\,J.\,A.~Baselmans, A.\,F.~Morpurgo, B.\,J.~van Wees, and T.\,M.~Klapwijk, Nature \textbf{397}, 43
(1999).

\bibitem{Dorokhov}
O.\,N.~Dorokhov, Solid State Commun. \textbf{51}, 381 (1984).

\bibitem{KO1}
I.\,O.~Kulik and A.\,N.~Omelyanchuk, Pis'ma Zh. Eksp. Teor. Fiz. \textbf{21}, 216 (1975) [JETP Lett.
\textbf{21}, 96 (1975)].

\bibitem{Brinkman}
A.~Brinkman and A.\,A.~Golubov, Phys. Rev.~B \textbf{61}, 11297 (2000).

\end{thebibliography}
\end{document}